\documentclass[twocolumn]{aastex63}

\submitjournal{ApJL}

\shorttitle{Black Hole Leftovers}
\shortauthors{Doctor, Farr, Holz}
\graphicspath{{./}{figures/}}

\usepackage{xcolor}

\newcommand{\numberdensity}{660_{-240}^{+440}\,\mathrm{Mpc}^{-3}}
\newcommand{\pctinnsc}{46_{-15}^{+17}}
\newcommand{\pctingc}{3_{-2}^{+3}}

\begin{document}

\title{
Black Hole Leftovers: The Remnant Population from Binary Black Hole Mergers\\
}

\correspondingauthor{Zoheyr Doctor}
\email{zoheyr.doctor@gmail.com}

\author[0000-0002-2077-4914]{Zoheyr Doctor}

\author[0000-0002-2916-9200]{Ben Farr}
\affiliation{Institute for Fundamental Science\\
The University of Oregon\\
1585 E 13th Ave\\
Eugene, OR 97403}

\author[0000-0002-0175-5064]{Daniel E. Holz}
\affiliation{Kavli Institute for Cosmological Physics, The University of Chicago, Chicago, IL, 60637, USA}
\affiliation{Enrico Fermi Institute, The University of Chicago, IL, 60637, USA}
\affiliation{Department of Physics, The University of Chicago, IL, 60637, USA}
\affiliation{Department of Astronomy \& Astrophysics, The University of Chicago, IL 60637, USA}

\begin{abstract}
The inspiral and merger of two black holes produces a remnant black hole with mass and spin determined by the properties of its parent black holes.  Using the inferred population properties of component black holes from the first two and a half observing runs of Advanced LIGO and Virgo, we calculate the population properties of the leftover remnant black holes. By integrating their rate of formation over the age of the universe, we estimate the number density of remnant black holes today.  Using simple prescriptions for the cosmic star formation rate and black hole inspiral delay times, we determine the number density of this leftover black hole population to be $\numberdensity$, corresponding to $\sim 60,000$ black hole remnants per Milky-Way-equivalent galaxy. The mass spectrum of these remnants starts at $\sim 10 M_\odot$ and can be approximated by a decreasing exponential with characteristic length $\sim 15 M_\odot$, the final spin distribution is sharply peaked at $\chi_f\sim0.7$, and the kick velocities range from tens to thousands of km/s. These kick velocities suggest that globular clusters and nuclear star clusters may retain up to $\pctingc\%$ and $\pctinnsc\%$ of their remnant black holes, respectively, while young star clusters would only retain a few tenths of a percent.
The estimates in this work assume that none of the remnants participate in subsequent hierarchical mergers.  If hierarchical mergers occur, the overall number density would drop accordingly and the remnant mass distribution shape would evolve over time. This population of leftover black holes is an inescapable result from gravitational-wave observations of binary black-hole mergers.
\end{abstract}



\section{Introduction} \label{sec:intro}
The Advanced LIGO \citep{AdvLigo} and Virgo \citep{AdvVirgo} gravitational-wave detectors have observed 46 binary black hole (BBH) mergers to date \citep{GWTC1, GWTC2}. Analysis of these gravitational wave (GW) sources yields estimates of the properties of the two component black holes comprising the binary before it merges (the ``inspiral" phase), as well as the properties of the single ``leftover'' remnant black hole. The properties of all the detected BBH mergers taken in tandem puts constraints on the intrinsic population distribution of mergers, which can help characterize the environments and circumstances under which BBHs form \citep[e.g.~][]{FishbachWhereBigBlackHoles,Taylor:2018iat,O3apop,Miller2020, Galaudage2020,Fasano2020, Tiwari:2020otp, Fishbach:2021yvy}.  Notably, the LIGO-Virgo Collaboration (LVC) has put constraints on the volumetric rates of binary black hole mergers as a function of black hole {\it inspiral} masses and spins, and finds that high mass mergers occur less often than low mass ones \citep{FishbachWhereBigBlackHoles,O3apop}, and that the spins of the black holes are not all perfectly aligned with their orbital angular momentum \citep{BFarrSpin,O3apop}.  These inferences have implications for the population of {\it remnant} black hole masses and spins, because the remnant properties are directly related to the inspiral properties \citep{Sedda:2018nxm,Sedda:2020vwo,GalvezGhersi:2020fvh}.  

A population of remnant black holes may have a range of observational consequences. For example, the remnants of BBH mergers could be recycled in subsequent ``hierarchical" mergers that are detectable through gravitational waves \citep{Fishbach:2017dwv,GerosaBerti,DoctorCoag,Kimball:2020opk}. Detecting a merger in which one or both of the BHs has parameters consistent with the remnant population but not with the ``1st generation" population could be a smoking gun of a hierarchical merger. For example, GW190521 could be formed hierarchically~\citep{2020PhRvL.125j1102A} (although see \citet{2020ApJ...904L..26F}).  The remnants could also lead to a range of other detectable effects, such as microlensing, X-ray emission from accretion, and/or interactions with other stars.  To facilitate these BH remnant studies, we characterize the distribution of masses, spins, and kick velocities of the remnant black holes using the inferred population of pre-merger black hole binaries.  

We emphasize that the LIGO/Virgo data robustly establish the existence of this leftover population of black holes, which are the final remains of the population of BBH mergers. As would be expected in the isolated formation scenario \citep[e.g.][]{Dominik:2012kk,Eldridge2017,Belczynski:2017gds,Mandel:2020cig}, we assume herein that these remnant black holes persist for the indefinite future. In particular, they do not undergo further mergers with other black holes. It is to be noted that some formation scenarios, such as dynamical formation, may lead to additional black hole interactions and mergers \citep{Fishbach:2017dwv,RodriguezNextGen,DoctorCoag,Kimball:2020opk,Kimball:2020qyd}. If these interactions occur with an appreciable rate, there would be subsequent evolution in the remnant BH population, altering the remnant distribution from that naively inferred through the merger rate-density distribution.

The remnant black holes have a number of characteristics in common with a purported population of primordial black holes \citep{Garcia-Bellido:2017fdg,Nishikawa:2017chy,DeLuca:2020qqa}. Recent interest in such a population has focused on the possibility of these black holes constituting the dark matter. There are a number of important distinctions between the remnant population which we discuss and primordial black holes. First, and most important: there is no doubt that a population of remnant black holes, such as we discuss, exists. They are an inevitable consequence of the mergers which LIGO/Virgo has observed. This population may suffer further consolidation into larger black holes, but no known physical processes can remove this remnant population from the universe\footnote{We neglect Hawking radiation, which in principle might erase this population of black holes on timescales greater than $10^{50}$ times the age of the universe.}. Primordial black holes remain speculative, and may not exist outside of theorist's imagination. A second important distinction is that the parent black holes of this remnant population are thought to form from stellar evolution and collapse. Their mass was therefore initially baryonic, and they therefore are not viable candidates for dark matter. Finally, as will be shown below, the expected mass to be found in this population is significantly less than the expected dark matter density.

In \S\ref{sec:methods} we describe our method for computing a posterior distribution on the remnant population properties given the inference of the pre-merger population. This method builds upon earlier work \citep{Fragione190521,FragioneRecoil} by using the LVC population inferences directly in the calculations.  \S\ref{sec:results} shows our resultant remnant population distribution based on GWTC-2, the latest LVC catalog of compact binary sources \citep{GWTC2}. We also discuss the potential imprints of these remnants and enumerate possible systematic errors. Finally, we offer concluding remarks in \S\ref{sec:conclusion}.

\section{Methods}\label{sec:methods}
The LIGO-Virgo Collaboration and other groups have made inferences of the population distribution of merging BHs \citep[e.g.][]{TalbotThrane,Taylor:2018iat,Fishbach:2019bbm,Roulet2020,O3apop,Kimball:2020qyd}.  These distributions are typically parameterized in terms of the {\it inspiral} parameters of the binaries, i.e.~the masses and spins of the component black holes prior to merger.  Since a given set of inspiral masses and spins uniquely determines the properties of the remnant through the theory of general relativity (GR), the inferences on the inspiral population imply inferences on the remnant population. 

The basis of our method is to post-process the inspiral population inferences to produce the remnant population. It should be mentioned that one could measure the remnant parameters for individual detected sources \citep[e.g.][]{Varma:2020nbm,GWTC2} and use them in an independent population analysis, but we opt for a post-processing method since it does not require parameterized remnant population models.  In the following subsections we describe the models, post-processing steps, and prescriptions used to perform our remnant population calculation.

\subsection{Incorporating existing population estimates}
The LVC has employed a suite of population models to describe the GWTC-2 dataset.  Herein, we use the LVC results from the population model with the highest Bayes factor, which is the Power-law + Peak model \citep{O3apop}.  In this model, the population of \textit{primary} masses (the more massive component in each binary) is parameterized as a mixture model of a power law plus Gaussian \citep{TalbotThrane}. The secondary mass is a power law conditioned on the mass of the primary. The spin magnitude and tilt distributions (at a reference GW frequency of 20 Hz\footnote{GWTC-2 parameter estimates reference all spins to 20 Hz except for GW190521, which was referenced to 11 Hz due to its large mass. Nevertheless, we treat the GW190521 spins as if they were referenced to 20 Hz. \citet{O3apop} has verified that this treatment results in errors smaller than the error between results with different gravitational waveform models.}) are assumed to be independent of one another and the masses. The spin magnitudes are described by a Beta distribution, and the cosines of the tilt angles with respect to the orbital angular momentum are parameterized with a mixture of a Gaussian and uniform distribution. Since the distribution of BH spin azimuthal angles (i.e.~spin direction in the orbital plane) are not expressly fit by the LVC population analyses, they inherit the prior used in LVC parameter estimation, namely isotropic in the orbital plane.  The full details of the model and priors can be found in \citet{O3apop}. The hyperparameters (e.g.~power law indices, Gaussian means and variances) that describe these distributions are constrained by the GWTC-2 data, and inferences on them are reported as ``hyperposterior" samples by the LVC.  Each hyperposterior sample $\Lambda_j$ implies a population distribution of inspiral parameters $p(\theta|\Lambda_{j})$, which can in turn be converted to remnant parameters.  

To incorporate these hyperposterior inferences, we start by generating a sample of inspiral masses and spins $\theta_i$ from a reference distribution $p(\theta|\Lambda_{\rm ref})$, where $\Lambda_{\rm ref}$ is the reference population model.  For each inspiral sample $\theta_i$ and each hyperposterior sample $\Lambda_j$, we calculate a weight $w_{ij}=p(\theta_i|\Lambda_j)/p(\theta_i|\Lambda_{\rm ref})$. We apply these weights to the \textit{remnant} parameter samples $\theta_i^{\rm rem}=f_{\rm GR}(\theta_i)$, which are calculated through a function $f_{\rm GR}(\theta)$ that is specified by GR.  These weights along with the remnant samples $\theta_i^{\rm rem}$ represent draws from the remnant distributions corresponding to each $\Lambda_j$.  Finally, to estimate and visualize the distributions of the remnants, we apply Gaussian kernel density estimates to the samples and weights, yielding continuous functions over the remnant parameter space. 

\subsection{Calculating Remnant Properties}
The mass, spin, and kick velocity of a BBH remnant are fully specified by the properties and configuration of the inspiraling BHs, and can be calculated in GR.  In practice, numerical relativity (NR) simulations are needed to compute the remnant parameters for an arbitrary set of inspiral masses and spins.  Since these simulations can be computationally expensive, fast surrogate models have been built to interpolate remnant parameters between NR simulations. We use two such surrogates here, namely the NRSur3dq8Remnant and NRSur7dq4Remnant models from the \texttt{surfinBH} package \citep{Varma:2018aht,Varma:2019csw}. We default to the NRSur7dq4Remnant model which calculates the final mass $M_f$, final spin vector $\vec{\chi}_f$, and kick velocity vector $\vec{v}_f$ given the 3D dimensionless spin vectors of the two inspiraling BHs and their mass ratio. The model is trained for mass ratios between 1 and 4 and can reliably extrapolate out to mass ratios of 6.  For mergers with $q>6$, we switch to the NRSur3dq8Remnant model that is trained out to $q=8$ but assumes the BH spins are aligned with the orbital angular momentum. In these $q>6$ cases, we ``ignore" the in-plane spin components by setting them to zero. While some systematic error is picked up from reverting to the NRSur3dq8Remnant model, we note that $q>6$ mergers are expected to be only a fraction of a percent of all mergers under the Power Law + Peak population. 

\begin{figure}[h!]
\epsscale{1.1}
\plotone{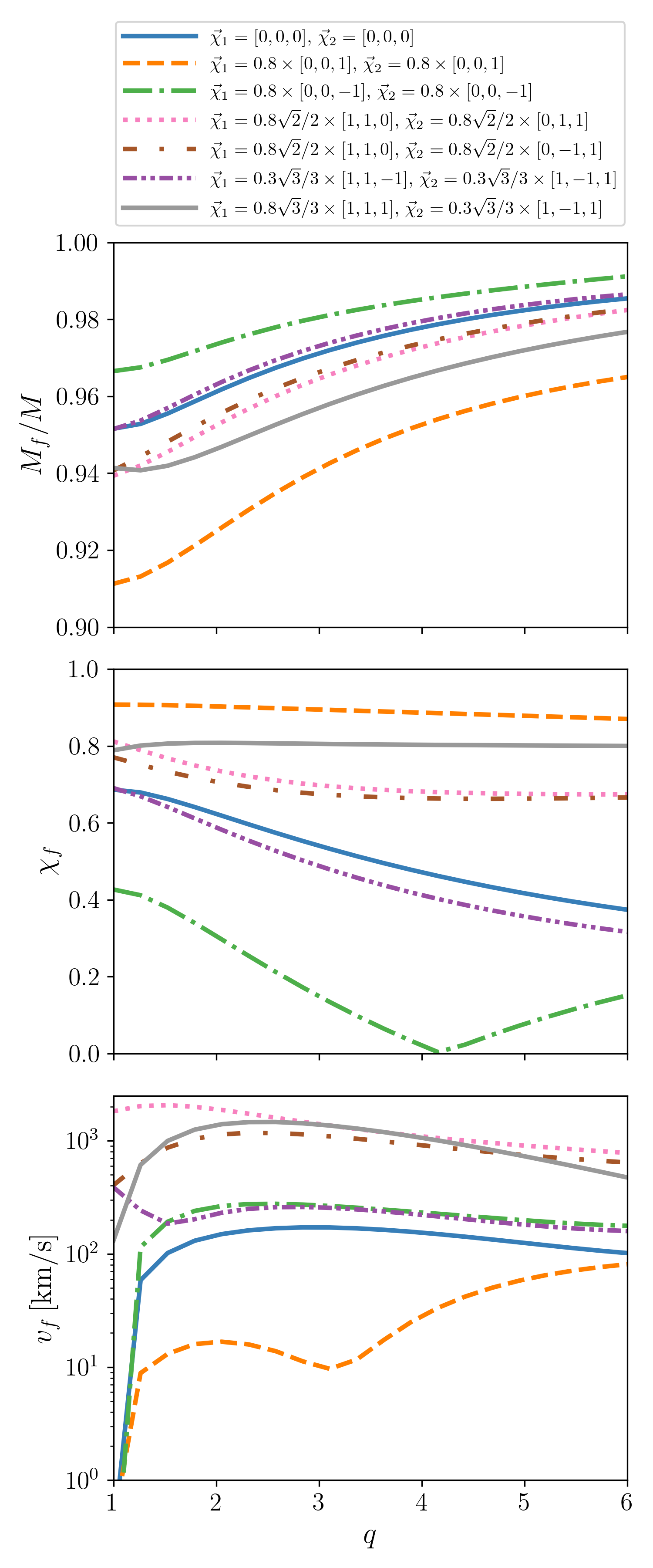}
\caption{Remnant parameters as a function of mass ratio for different fixed spin configurations.  Each line style corresponds to a different spin configuration in the legend.  $\vec{\chi}_1$ and $\vec{\chi}_2$ are the spins of the inspiraling black holes defined at $t=-100M$ before the peak of the GW in the co-orbital frame, which is the default convention used in NRSur7dq4Remnant. $M$ is the inspiral total mass. \label{fig:eg_params}}
\end{figure}

To show how the BH remnant parameters change with binary inspiral parameters, we display the remnant properties calculated via NRSur7dq4Remnant for a set of example cases in Figure \ref{fig:eg_params}. Each line shows the remnant parameters as a function of mass ratio for a fixed spin configuration (defined at $t=100M$ before peak of the GW). In the top panel, we see that the ratio of final mass $M_f$ to initial mass $M$ follows a common trend regardless of spin configuration: it is greater than 0.9 and approaches 1 as the mass ratio increases. On the other hand, the final spin and final velocity are quite sensitive to spins and mass ratio \citep{Fishbach:2017dwv}. Although $\chi_f \sim 0.7$ at $q=1$ for most cases, there can be significant deviations from that depending on the spins and mass ratio. For example, when the inspiral spins are large and anti-aligned with the orbital angular momentum (green, dash-dotted), the final spin decreases to near zero with increasing mass ratio until $q=4$ and then increases. The final velocity is the most sensitive to input configuration, though the kick velocities tend to be highest with large, misaligned component spin magnitudes.   

One other feature of these surrogate models is that they are capable of reporting 1-$\sigma$ uncertainties on remnant parameters due to interpolation error \citep{Varma:2018aht,Varma:2019csw}.  To account for the surrogate model uncertainty in our population analysis, we dither each calculated remnant parameter using the surrogates' reported 1-$\sigma$ intervals. 

\begin{figure*}[ht!]
\epsscale{1.1}
\plotone{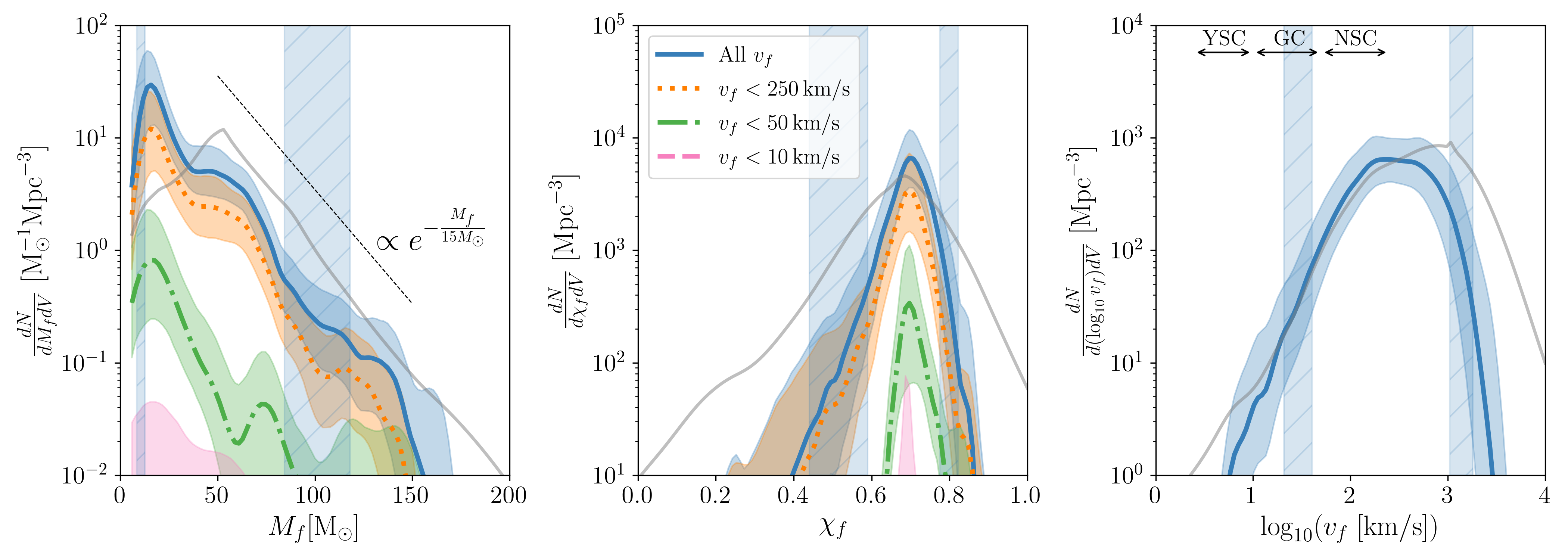}
\caption{Inferred mass $M_f$ (left), spin magnitude $\chi_f$ (middle), and kick velocity $v_f$ (right) distributions of remnant black holes. Each panel shows the number density spectrum $dN/(dVd\theta^{\rm rem})$ versus remnant parameter $\theta^{\rm rem}$. The solid blue lines show the medians of multiple number density spectrum samples which are each constructed with a Gaussian kernel density estimate of the population-weighted remnant parameter samples.  The light blue regions around the blue lines show the 90\% symmetric confidence intervals on the number density spectrum values. Gray lines denote expected remnant distribution from the population hyper-prior, up to arbitrary normalization. In the left and middle panels we also show the number density distributions of remnants below different kick velocity thresholds. We consider thresholds of 10 km/s (pink, dashed), 50 km/s (green, dashed-dotted), and 250 km/s (orange, dotted), which correspond roughly to the upper ends of the ranges of escape velocities for young star clusters, globular clusters, and nuclear star clusters, respectively. At the top of the right panel, the approximate ranges of escape velocities of young star clusters (YSC), globular clusters (GC), and nuclear star clusters (NSC) are shown with the arrows.  The vertical hatched bands show the 90\% symmetric intervals on the 1st and 99th percentiles of the inferred remnant population distributions. \label{fig:1D_dists}}
\end{figure*}

\subsection{Calculating the number density of remnants}
In addition to the shape of the population distributions of mergers, the LVC has inferred the total number of mergers occurring per comoving time and volume in the local universe \citep{O3apop}.  Integrating this rate density from the beginning of the universe to now yields the local number density of remnants. One complication in this number density calculation is that the rate density of mergers is likely not constant over cosmic time, because the star formation rate evolves over time, and BBH mergers are expected to occur with some delay from the initial formation of the progenitor stars. To incorporate these considerations, we use a procedure analogous to that in \citet{LVCKN}. We convolve the star formation rate from \cite{Madau:2016jbv} with a $t^{-1}$ distribution of delay times from progenitor formation to BBH merger and a minimum delay time of $t_{\rm min} = 10$ Myr. We then assume this convolution is proportional to the merger rate density $R$.  That is, $R(t) = A \int_{t_{\rm min}}^{t_h} p(t-t')\psi(t')dt'$, where $A$ is a normalization constant, $p(t-t') = (t-t')^{-1}$, $t_h$ is the Hubble time, and $\psi(t)$ is the cosmic star formation rate. Integrating this over cosmic time yields 
\begin{equation}
    n = R_0 \frac{\int_0^{t_h}\int_0^t p(t-t')\psi(t')dt' dt}{\int_0^t p(t-t')\psi(t')dt'}
\end{equation}
where $R_0$ is the rate density today, which we take to be the local rate inferred by the LVC. For each hyperposterior sample $R_0$ we compute $n$ to build a posterior distribution on the number density.  

\section{Results and Discussion}\label{sec:results}

\begin{figure*}[ht!]
\epsscale{1.2}
\plotone{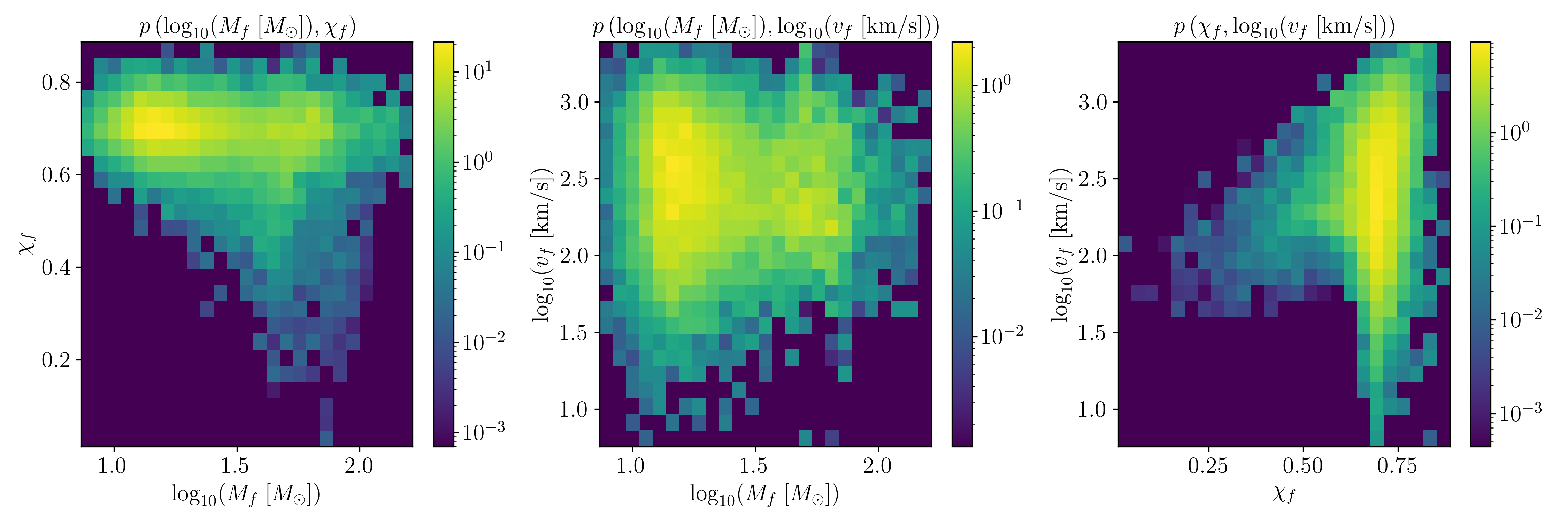}
\caption{2D joint posterior population distributions on expected remnant parameters.  The color shows the probability of a given pair of remnant parameters, marginalized over all inspiral population hyperparameter samples.  \label{fig:2D_dists}}
\end{figure*}

Using the calculated weights, remnant parameters, and number densities as described in \S\ref{sec:methods}, we show our inferences on the remnant population in Figures \ref{fig:1D_dists} and~\ref{fig:2D_dists}. At 90\% credibility, we find the present-day number density of BBH remnants to be $n = \numberdensity$. Figure \ref{fig:1D_dists} shows the 1D number density distributions of final masses, spins, and kick velocities of the remnant population. The y-axes represent $dN/(dVd\theta^{\rm rem})$, the number of remnants $N$ occurring in a comoving volume $V$ per unit remnant parameter $\theta^{\rm rem}$. The solid blue lines are the median number density distribution estimated from our population-weighted remnant parameters and kernel density estimates. The surrounding light blue bands show symmetric 90\% confidence intervals for the number density spectrum at each remnant parameter value. For reference, we also plot the expected remnant distribution from the population hyper-prior, up to arbitrary normalization, in gray in each panel. Lastly, the vertical hatched bands denote the symmetric 90\% intervals for the 1st and 99th percentiles of the distributions. We find that the remnant masses follow an approximate declining exponential with characteristic length $15M_\odot$, the final spin distribution is tightly peaked at $\chi\sim 0.7$, and there is a wide range of kick velocities from tens to thousands of km/s. In all cases, the inferred distributions differ from the expected distributions from the population hyper-priors shown with gray lines, meaning that the GWTC-2 detections are informative with regards to the remnant population. Notably, the mass distribution has a bump-like feature around 40--70$M_\odot$ corresponding to components from the Gaussian peak in the Power-Law + Peak model. Additionally, the right-most panel annotations show the approximate range of escape velocities for young star clusters (YSCs), globular clusters (GCs) and nuclear star clusters (NSCs), and we see that many remnants could be retained in a cluster environment. 

To further investigate the population of remnants that could be retained in star clusters, we re-evaluate our final mass and final spin number density distributions but only including remnants with kick speeds less than an ejection threshold $v^*$. We consider $v^* = [10, 50, 250]\,\mathrm{km/s}$, and plot the resultant distributions in Figure \ref{fig:1D_dists} in pink (dashed), green (dashed-dotted), and orange (dotted), respectively.  These thresholds roughly bound the escape velocities of YSCs, GCs, and NSCs \citep{Mapelli:2020xeq}. The small escape velocities of YSCs allow very few remnants to stick around, which is why the pink distributions barely enter the bottom of the plots. We estimate that no more than a few tenths of a percent of all remnants created could be retained in a YSC. In contrast, GCs and NSCs have escape velocities that would allow $\pctingc\%$ and $\pctinnsc\%$, respectively, of our inferred remnant population to be retained.   

To visualize the correlations between remnant parameters $\theta_i$ and $\theta_j$, we plot their 2D posterior population distributions $p(\theta_i,\theta_j) = \int d\Lambda p(\theta_i,\theta_j|\Lambda) p(\Lambda|\textrm{GWTC-2})$ in Figure \ref{fig:2D_dists}. Most strikingly, the final spin distribution has support for smaller values only when the final mass is large.  Since the Power Law + Peak BH mass distribution cuts off at low BH masses, the mass ratio---to which the final spin is quite sensitive---can only become large for high primary mass. Likewise, unequal mass ratios can result in large kicks as seen in Figure \ref{fig:eg_params}, so remnants that experience the smallest kicks come from near-equal-mass mergers and therefore have final spins exclusively around 0.7.  This can also be seen in the middle panel of Figure \ref{fig:1D_dists}: the green final spin distribution for $v_f<50\,\,\mathrm{km/s}$ does not have support at low spins, unlike the cases that allow higher escape velocities.

\subsection{Prospects for BBH Remnant Detection}\label{subsec:propsects}
As mentioned in Section \ref{sec:intro}, there are two primary ways one might detect a BBH remnant: through gravitational waves emitted by hierarchical mergers or though ``indirect" electromagnetic detections of BHs. The existence of hierarchical mergers is still uncertain, but the eventual reach of future gravitational-wave detectors could probe these if they are occurring with an appreciable rate \citep{Hall2019,Ng:2020qpk}.  If hierarchical mergers are rare, a BBH detected with a component coming exclusively from the distributions in Figure \ref{fig:1D_dists} could be a smoking gun of a hierarchical merger with a BBH remnant.

On the other hand, if hierarchical mergers are common, then the GWTC-2 data set and LVC population inferences may already include hierarchical mergers. In that case, our results cannot be interpreted as present-day number densities of remnants, because we do not account for remnants merging again. Instead, our resultant number density distributions could be re-normalized to the inferred {\it merger} rates and interpreted as the distributions over parameters of remnants that are being {\it created} per unit time-volume. Such a re-normalization would make our results directly analogous to the inferences of the inspiraling population: they are inferences on numbers of events per volume {\it and time}, not a number density.   
To properly compute the present day number density of remnants in this strongly hierarchical scenario, we would need to include second generation mergers through a self-consistent hierarchical population model \citep{Kimball:2020opk,DoctorCoag}. There are tantalizing hints that some events in GWTC-2 may have a hierarchical origin \citep{Kimball:2020qyd,2020arXiv201110057Z}, so future work could include this possibility in the number density calculation.  

Now we turn to electromagnetic means of detecting BBH remnants. As with hierarchical mergers, the detectability of electromagnetic signals from BBH remnants is contingent on the number of remnants, the remnants' likelihood of interacting with other bodies, and the statistical power to differentiate a remnant signal from a non-remnant BH signal background. With a local remnant number density of $\numberdensity$, one could expect $O(60,000)$ remnants produced in the Milky Way\footnote{This assumes a conversion factor of $\sim$ 0.01 Milky-Way-equivalent galaxies per cubic Mpc, which is similar to the conversion used in \citet{Abadie:2010cf}.}, which matches theoretical expectations \citep[e.g.][]{Lamberts2018,Olejak:2019pln}. This number of remnants, even ignoring that some may be ejected from the galaxy due to kicks, is far below the limits set on primordial black hole number densities by microlensing studies, which probe populations of black holes comprising percents of the total dark matter mass \citep{Lu:2019hoc,Tisserand:2006zx,Garcia-Bellido:2017fdg}. As such, we do not expect that these remnants can be detected with current instruments with microlensing. Another avenue for detection might be through observations of dynamics of companion stars or X-ray emission from accretion onto a remnant.  These detections are possible only under the highly speculative condition that the remnants encounter other bodies after their creation, which could occur in a cluster environment or if the remnant approaches the center of the galaxy due to dynamical friction. The results of our study could serve as a guide for extracting whether BHs detected via electromagnetic means come from a remnant or stellar population.  

\subsection{Sources of Systematic Error}
In this work we have made a number of assumptions which could affect the resultant remnant parameter distributions and number densities.  Firstly, we only consider the results of the LVC's inferences with the Power Law + Peak model.  If the inspiral distribution has more complicated features that are not resolved by this model, the resultant remnant parameter distributions may also pick up additional features. However, the LVC's analyses with other models show a consistent picture of a lower rate of mergers with larger mass, meaning the overall shape of the remnant mass distributions in Figure \ref{fig:1D_dists} should be relatively insensitive to other model choices. Likewise, the final spins of mergers tend to be near $\chi_f=0.7$ except in select cases, so the spin distribution is also unlikely to change dramatically with different parameterizations.  On the other hand, the kick velocities are much more sensitive to the assumed models. Changes to the parameterization of inspiral mass ratio or spins could affect the kick velocities.  For example, if the component BH spin angles in the orbital plane do not match our assumption of being uniformly distributed, which could happen in some dynamical environments \citep{Mould:2020cgc,Yu:2020iqj}, certain kick speeds could become \mbox{(dis-)favored}. Changes in spin orientations alone can change the kick velocities significantly, as seen in Figure~\ref{fig:eg_params}. 

In calculating the number density, we assume that the rate of black hole mergers follows the star formation rate of \citet{Madau:2016jbv} with a $t^{-1}$ delay time and minimum delay time of 10 Myr. Since the star formation history and black hole merger delay times are still relatively uncertain, the overall number density of remnants calculated herein is subject to change with stronger constraints, but unlikely to change by the multiple orders of magnitude needed to affect our conclusions about remnant detectability, assuming the binaries come from a stellar population.  For example, if we use the rate density versus redshift profile from \citet{Rodriguez:2018rmd} for BBHs from globular clusters and normalize to the inferred LVC rate at $z=0$, the inferred present-day number density changes by $< 10\%$. We neglect the possibility that the binaries come from primordial black holes, which would entail a different redshift evolution of the merger rate. 

\section{Conclusion}\label{sec:conclusion}
When black holes merge, they inevitably leave behind a single leftover black hole. The population of black hole remnants produced in a given time and volume is set by the population of inspiraling and merging black holes, because general relativity predicts the final state of black hole mergers.  Assuming the remnants persist, the rate of mergers can be integrated over cosmic time to yield a present-day number density of black hole remnants.  Using the population of inspiraling black holes inferred by the LIGO-Virgo Collaboration and a simple prescription for the cosmic BBH merger rate, we find the present day number density of black holes to be $\numberdensity$. 

We incorporate a surrogate model for calculating black hole remnant properties to determine the spectrum of BH remnant properties.  These spectra show that remnant masses are distributed roughly as a decreasing exponential with length scale $\sim 15M_\odot$, and the remnant spins are centered near $\chi_f \sim 0.7$. There is a wide range of kick velocities expected for these remnants, with about half kicked with speeds over 250 km/s. A few percent of these remnants could be retained in globular clusters and up to half could be retained in nuclear star clusters.

In principle, the remnant population could be directly measured through gravitational waves from hierarchical mergers or through electromagnetic observations of galactic systems.  However, the low number density of these remnants poses a significant observational challenge, especially for techniques such as microlensing. Nevertheless, the results herein lay the groundwork for future remnant black hole searches.  The inferred distributions and software associated with this analysis are available at \url{https://github.com/zodoctor/final_state_population}.

\acknowledgments{
Z.D.~would like to thank Chase Kimball, Vijay Varma, David Keitel, Christopher Berry, Nathan Johnson-McDaniel, and Ilya Mandel for helpful conversations and comments during preparation of this manuscript. B.F.~is supported by NSF grant PHY-1807046. D.E.H.~is supported at
the University of Chicago by the Kavli Institute for
Cosmological Physics through an endowment from the
Kavli Foundation and its founder Fred Kavli. D.E.H.~is also supported by NSF grant PHY-2011997, and gratefully acknowledges the Marion and Stuart Rice Award.
}

\bibliography{references}{}
\bibliographystyle{aasjournal}



\end{document}